\documentclass{rmaa}
\def\spose#1{\hbox to 0pt{#1\hss}}
\def\lta{\mathrel{\spose{\lower 3pt\hbox{$\mathchar"218$}}
     \raise 2.0pt\hbox{$\mathchar"13C$}}}
\def\gta{\mathrel{\spose{\lower 3pt\hbox{$\mathchar"218$}}
     \raise 2.0pt\hbox{$\mathchar"13E$}}}
\title{Preservation of Cuspy Profiles in Disk Galaxy Mergers}
\author{H\'ector Aceves\altaffilmark{1} \& H\'ector Vel\'azquez
  \affil{Instituto de Astronom\'{\i}a, UNAM, Ensenada}
  }
\altaffiltext{1}{Email: \texttt{aceves@astrosen.unam.mx}}
\fulladdresses{
\item H\'ector Aceves \& H\'ector Vel\'azquez:
  Instituto de Astronom\'{\i}a, UNAM Campus Ensenada,
  Apdo Postal 877, Ensenada 22800, Baja California, M\'exico
  (aceves@astrosen.unam.mx, hmv@astrosen.unam.mx)}

\shortauthor{Aceves \& Vel\'azquez} \shorttitle{Preservation of Cuspy Profiles}
\SetVolume{00} \SetFirstPage{000} \SetYear{2004}
\ReceivedDate{------}
\AcceptedDate{------}
\SetYear{2005}
\resumen{Se efectuaron tres simulaciones auto-consistentes de 
$N$-cuerpos de fusiones entre galaxias, con un perfil
pronunciado de materia oscura, para estudiar si la pendiente interna
del perfil de densidad oscura se preserva en los remanentes.  
En estas simulaciones los progenitores incluyen un disco estelar y un 
esp\'{\i}n intr\'{\i}nsico en el halo a diferencia de estudios similares.
 La raz\'on de masas de las galaxias progenitoras 
es de alrededor de 1:1, 1:3 and 1:10. Encontramos que el perfil de densidad 
inicial pronunciado de los halos oscuros se preserva en los remanentes
para los casos considerados aqu\'{\i}.
}

\abstract{We carried out three self-consistent $N$-body simulations 
of galaxy mergers, with a cuspy dark matter profile,
in order to study if the inner dark density slope is preserved 
in the remnants. In 
these simulations the progenitors include both a stellar 
disk and an intrinsic angular momentum for the halos, unlike previous 
similar studies. 
The mass-ratios of progenitor galaxies are about 1:1, 1:3 and 1:10. 
We find that the 
initial cuspy density profile of the dark halos is preserved in the
remnants for the cases considered here.
}

\keywords{
galaxies: kinematics and dynamics -- galaxies: dark matter -- 
methods: $N$-body simulations.
}

\begin{document}
\maketitle
%%%%%%%%%%%%%%%%%%%%%%%%%%%%%%%%%%%%%%%%%%%%%%%%%%%%%%%%%%%%%%%%%%%%%%%%%
%%%%%%%%%%%%%%%%%%%%%%%%%%%%%%%%%%%%%%%%%%%%%%%%%%%%%%%%%%%%%%%%%%%%%%%%%
%\baselineskip=20pt

%%%%%%%%%%%%%%%%%%%%%%%%%%
\section{Introduction}
%%%%%%%%%%%%%%%%%%%%%%%%%

High-resolution simulations of hierarchical structure formation  
within the $\Lambda$CDM cosmogony show that the inner regions of the 
halos follow a power-law cusp, $\rho \propto r^{-\gamma}$, with   
$1 \leq \gamma \leq 1.5$ while in their outer parts the profiles go as  
$\rho\propto r^{-3}$ over a wide range of mass-scales (e.g., 
Navarro, Frenk \& White 1997, hereafter NFW, Moore et~al.~1999,
hereafter M99, Fukushige, Kawai \& Makino 2004, Navarro et~al.~2004).

The physical origin of this profile is still unclear and it is the 
subject of an intense investigation. Nevertheless, there is a certain 
agreement in the sense that such density profile is linked to the 
accretion and merging history of the dark matter substructures (e.g., 
Syer \& White 1998, Dekel, Devor \& Hetzroni 2003, Williams, Babul \& 
Dalcanton 2004).

Several authors have studied the shape and survival of the density profile
inner slope in collisionless
mergers (e.g., White 1978, Fulton \& Barnes 2001, Boylan-Kolchin \& Ma 2004).
In the early study of White (1978) it was found that mergers of cored
galaxies were more concentrated than their progenitors. Fulton \& Barnes (2001)
found in their $N$-body experiments that steep cusps are preserved during 
mergers of equal-mass progenitors. Boylan-Kolchin \& Ma (2004, hereafter BKM)
concluded that if two galaxies merge having each one a 
core-type profile the resulting remnant will be core-type, and if one 
progenitor has a cuspy profile then the remnant will also be cuspy. 
A result in agreement with other simulations (Moore et~al.~2004).

   These type of studies find justification in that although in a collisionless
simulation the phase-space density must be conserved, due to Liouville's 
theorem, there is no a priori reason to expect that both 
configuration and momentum distributions are to be conserved independently 
(BKM).

A common feature of previous related works is that they consider 
gravitational systems with only a spherical or dark matter component,
and they do not consider the effect, for example, that 
a disk or an intrinsic halo spin might have on the preservation of cusps 
during a merger process. This needs to be addressed since there are
several studies that indicate that angular momentum
may play a key role at the time of formation of the halo as well as on 
 the shape of the density profile within the inner regions, by preventing that 
dark matter particles reach the inner regions, resulting in a shallower 
central profile (e.g., White \& Zaritsky 1992, Hiotelis 2002, 
Ascasibar et~al.~2004). Furthermore, cosmological simulations lead to
dark halos having an intrinsic angular momentum (Barnes \& Efstathiou 1987,
 Lemson \& Kauffmann 1999).

In this work we study three numerical $N$-body simulations of progenitor 
self-consistent disk galaxies with mass-ratios of about 1:1, 1:3 and 1:10, 
 following parabolic encounters. 
The present work expands on earlier similar studies mainly in that the  
progenitors have a disk component and  their halos have 
an intrinsic angular momentum. 
This would allows us to test, using more realistic initial conditions,
 if the cuspy nature of the progenitor 
galaxies is preserved in mergers; at the resolution provided by our 
$N$-body simulations.

The rest of this paper has been organized as follows. 
In $\S$2 we describe the theoretical method used to construct the
galaxy models, the properties of their dark halos, especially the behavior 
of the inner logarithmic derivative
 and the initial conditions for the encounters. 
In $\S$3 we present our results, and in 
 $\S$4 we summarize our main results.

%%%%%%%%%%%%%%%%%%%%%
\section[]{Merger Model}
%%%%%%%%%%%%%%%%%%%%%

\subsection{Disk Galaxies}
%%%%%%%%%%%%

Our galaxy models consist of a spherical halo and a stellar disk; no bulge 
component is included. The disk is represented by an exponential 
profile in the radial direction and by an isothermal sheet in the vertical one 
(Hernquist 1993):

\begin{equation}
\rho_{\rm d} (R,z) = \frac{M_{\rm d}}{4 \pi R^2_{\rm d} z_d} \exp( -
R/R_{\rm d})  \, {\rm sech}^{2} (z/z_d) \;,
\end{equation}
where $R_{\rm d}$ and $z_{\rm d}$ are the radial and vertical scale-lengths 
of the disk, respectively. 

For the dark halo, a spherical model is adopted by assuming a NFW-profile 
with an exponential cutoff:
\begin{equation}\label{eq:nfwm}
\rho_{\rm h} (r) = \frac{M_{\rm h}\, \alpha_{\rm h} }{4 \pi r  (r + r_{\rm 
s})^2} \ \exp\left[ - \left(\frac{r}{r_{200}} + q \right)^2  \right]  \;,
\end{equation}
where $r_{\rm s}$ and $r_{200}$ are the the scale and ``virial'' radii, 
respectively, $c\!=\!q^{-1}\!=\!r_{200}/r_{\rm s}$ and $M_{\rm h}$ are the 
concentration and the mass of the halo.   The $r_{200}$ is defined in
cosmological simulations as the radius where the mean interior density is 200
times the critical density. In general, $r_{200}$  is not  exactly the
virial radius ($\S2.4$), but is a good approximation to it and serves to
characterize the halo (NFW, Cole \& Lacey 1996).\footnote{\footnotesize The 
virial radii of the disk numerical models in $\S2.4$ 
turnout to be $\approx 20$\% smaller than the $r_{200}$ in Table 1.}
 Here, 
$\alpha_{\rm h}$ is a normalization constant given by
$$
\alpha_{\rm h} = \frac{\exp( q^2) }{ \sqrt{\pi} q \exp(q^2) {\rm Erfc}(q) + 
\frac{1}{2} \exp(q^2) {\rm E}_1(q^2) - 1 }
$$
being ${\rm Erfc}(x)$ the complimentary error function and ${\rm E}_1(x)$ 
the exponential integral.

The radial scale length of the disk is obtained accordingly to the Mo, 
Mao \& White (1998) framework of disk galaxy formation. To determine 
this quantity five parameters 
are required, namely: the circular velocity $V_{\rm c}$ at 
$r_{200}$, the dimensionless
spin parameter $\lambda$,  the concentration $c$ of the dark halo, 
 the fraction of disk to halo mass $m_{\rm d}$,  and the
fraction of angular momentum in the disk to that in the halo $j_{\rm d}$.
We followed the method described by Shen, Mo \& Shu (2002) to obtain these
parameters.  
A $\Lambda$CDM cosmology is adopted with a mass density of $\Omega_0=0.3$  
and with a contribution due to the cosmological constant of 
$\Omega_\Lambda=0.7$.

The selected galaxy disks satisfy a  stability criterion 
$\varepsilon_{\rm m}$$\,\ge\,$$0.9$; where
 $\varepsilon_{\rm m}$=$V_{\rm m}
(G M_{\rm d} / R_{\rm d})^{-1/2}$, being $V_{\rm m}$  the maximum rotation
velocity (Efsthatiou, Lake \& Negroponte 1982, Syer, Mao \& Mo 1997). 
Mass-ratios of about 1:1, 1:3 and 1:10 were chosen to study the properties 
of the merger remnants. Finally, velocities of particles 
were set up through Jeans equations following the method prescribed 
by Hernquist (1993).

In Table~\ref{tab:glxmodels} we list the parameters  defining our
galaxy models; below the merger simulation  label ($S1$, $S2$, and $S3$)
 the mass ratio of the secondary to the primary galaxy has been indicated.
The number of particles in the halo $N_{\rm h}$ and in the disk $N_{\rm d}$ 
are also listed.

%%%%%%%%%%%%%%%%%%%%%%%%%
\begin{table*}
 \centering
  \caption{Initial Galaxies}\label{tab:glxmodels}
  \begin{tabular}{c|crccrr|cccr}  \hline\hline
   Merger    & \multicolumn{6}{c}{Halo Properties} & 
             \multicolumn{4}{c}{Disc Properties} \\ 
  $M_{\rm s}/M_{\rm p}$  & $M_\mathrm{h}$ & $r_\mathrm{200}$ & $\lambda$ & $j$
     & $c$ & $N_{\rm h}$  & $M_\mathrm{d}$ & $R_\mathrm{d}$ & $z_\mathrm{d}$ &
 $N_{\rm d}$ \\
    & [M$_\odot$] & [kpc] &  & &  & & [M$_{\odot}$] & [kpc] & [kpc] &  \\
 \hline
 $S1$ & $4.74\times 10^{11}$ & 110.3 & 0.112 & 0.106 & 11.39 & 240000 & $4.16\times 10^{10}$ & 6.6 & 1.29 & 60000  \\
0.14  & $6.99\times 10^{10}$ & 58.3 & 0.071 & 0.035 & 9.94 & 165992 & $2.88\times 10^{9}$ & 1.6 & 0.19 & 48689 \\ \hline
 $S2$  & $4.02\times 10^{11}$ & 104.3 & 0.063 & 0.041 & 7.63 & 57126 & $2.11\times 10^{10}$ & 2.4 & 0.39 &  12000 \\
  0.32  & $1.25\times 10^{12}$ & 152.3 & 0.056 & 0.068 & 3.84 & 177571 & $7.78\times 10^{10}$ & 5.7 & 0.47 & 44267  \\
\hline
 $S3$ & $8.11\times 10^{10}$ & 61.2 & 0.099 & 0.142 & 10.87 & 150000 & $7.22\times 10^{9}$ & 4.6 & 0.73 & 30000  \\
 0.98  & $8.33\times 10^{10}$ & 61.8 & 0.122 & 0.095 & 10.01 & 154050 & $7.15\times 10^{9}$ & 3.9 & 0.56 & 29686 \\
\hline\hline
\end{tabular} 
\end{table*}
%%%%%%%%%%%%%%%%%%%%%%%%%%%%%%%

\subsection{Encounter Parameters}
%%%%%%%%%%%%%

Only parabolic encounters are considered for the binary mergers. The relative 
initial separation between both galaxy centers is taken to be
\begin{equation}
R_{\rm s} = 1.25 \, ( r_{200,1} + r_{200,2} );
\end{equation}
where $r_{200,1}$ and $ r_{200,2}$ are, respectively, their $r_{200}$ 
``virial'' radii.

Orbital angular momentum is introduced in the simulations by 
randomly choosing a pericenter for each encounter, assuming that galaxies 
are point particles in a Keplerian orbit. Pericentric radii in the range
$R_{\rm p} =\{5-20\}$~kpc were adopted. These values for the orbital 
parameters are consistent with those found in cosmological 
$N$-body simulations and tend to favor mergers 
(Navarro, Frenk \& White 1995). The relative orientation of the spin 
vector of the galaxies, relative to the orbital plane of the encounter, 
is chosen randomly.

\subsection{Numerical Tools}
%%%%%%%%%%%%%

These $N$-body simulations were performed using a parallel version of {\sc
  GADGET}, a tree base code with individual timesteps 
(Springel, Yoshida \& White 2001). The runs
  were made in a Pentium based cluster of 32 processors (Vel\'azquez \& 
Aguilar  2002). Softening parameters $\epsilon_{\rm d} \!=\! 35\,$pc 
and $\epsilon_{\rm h}\!=\!350\,$pc for disk and halo particles,  
respectively, were used. Since
{\sc GADGET} uses a spline kernel for the softening, the gravitational
interaction between two particles is fully  Newtonian for separations 
larger than twice the softening parameters \citep{P03}.

The typical time of  arrival of the galaxies to the first passage through
 pericenter is about $1\,$Gyr. We follow each encounter for a total 
time of about $8\,$Gyr. At this time the remnants had reached a 
stable value close to virial 
equilibrium. Energy conservation was better than 0.25\% in all our 
simulations.

\subsection{Resolution Criteria and Stability of Initial
 Dark Halo Profiles}
%%%%%%%%%%%%%%%%%%%%%%%%%%%%%%%%%%%%%%%%%%%%%%%%%%%%%%%

The two-body relaxation time-scale imposes an inner `convergence' 
radius $r_{\rm c}$ over which the stellar system can be adequately 
described by a collisionless distribution function (e.g., Power et~al.~2003; 
Hayashi~et~al.~2004; Diemand~et~al.~2004; Binney~2004).

Here, we take $r_{\rm c}$ as the radius where
 its local two-body relaxation time-scale $t_{\rm r}$ is  
equal to the period 
$T_{\rm v}$ of a circular orbit at the virial radius; i.e., $r_{\rm c}$
satisfies 
\begin{equation}\label{eq:rconv}
\frac{t_{\rm r}(r)}{T_{\rm v}} = \frac{N}{8 \ln N} \frac{r}{V(r)} 
\frac{V_{\rm v}}{r_{\rm v}}=1\;,
\end{equation}
where $V_{\rm v}$ the circular velocity at the virial radius, $r_{\rm v}$, 
and $N \! = \! N(r)$ the 
total number of particles inside a radius $r$. The virial radius is 
 $r_{\rm v}\!\! = \!\! G M_{\rm t}/|W|$, being $M_{\rm t}$ the total
 mass of all bound particles and $W$ its total potential energy. 

We computed spherically averaged density profiles, $\rho(r)$, for each
initial galaxy in a logarithmically spaced grid within $r_{\rm c}$ and
  $r_{\rm v}$; in general, we find $r_{\rm c}/r_{\rm v} \!\approx \! 0.01$. 
To check that the dark density profile of our initial galaxy models does not
change, we evolved in isolation each galaxy for about 8~Gyr, that corresponds
to $\approx 25$ dynamical times, 
$t_{\rm{dyn}}=\sqrt{3 \pi/16 G {\bar \rho}_{\rm h}}$;
where ${\bar \rho}_{\rm h}$ is the mean density at the half-mass
radius. At this time 
the virial ratio fluctuates within $\lta 1$\% from the ideal value
of unity, and no significant 
change in the density profiles was observed; see Figure~1. 

%%%%%%%%%
\begin{figure}
\centering
\includegraphics[width=8.3cm]{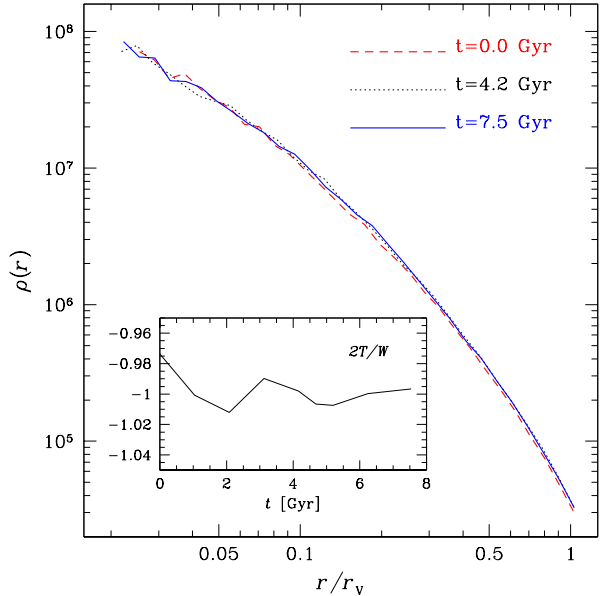}
\vspace{-0.7cm}
\caption{Density profiles of the dark matter component at different times 
of a galaxy model evolved in isolation. Different lines indicate the time
spent in isolation. The dynamical time is $t_{\rm dyn}\approx 0.3\,$Gyr, 
so the evolution is for $\approx 25 t_{\rm dyn}$. The inner profile, 
close to $r_{\rm c}$, 
remains stable showing no shallowing
of the density slope. The inset indicates the evolution of the virial
ratio.
}
\label{fig:isoevol}
\end{figure}
%%%%%%%%%%%

Results in the literature (e.g.,~Kazantzidis, Ma-gorrian \& Moore
2004, Springel et~al.~2005) indicate that obtaining the velocities 
of particles using the Hernquist (1993) procedure, that assumes
a Maxwellian local velocity distribution,  is inadequate.
In these works it is found that  models evolved in isolation
 tend to relax rapidly to a shallower 
inner density slope than the initial condition. 
However, our models constructed using Hernquist's method do not
show a shallowing of the inner slope when evolved in isolation 
(see Figure~1).

We fitted NFW density profiles,
\begin{equation}\label{eq:nfw}
\rho_{*}(r) = \frac{ \rho_{\rm s} }{ (r/{\bar r}_{\rm s}) (1 + r/{\bar 
r}_{\rm s})^2} \;,
\end{equation}
to the $\rho(r)$ of the initial numerical models; where ${\bar r}_{\rm s}$ 
is the scale radius.
The fitting was done by $\chi^2$--minimization using the Levenberg-Marquardt
 method \citep{NR}. From the fitted profile we computed
 the logarithmic derivative $\beta(r)={\rm d}\log \rho/{\rm d}\log r$.

To verify that our binning method does not affect the fitting of the density 
profiles, we computed $\rho (r)$ in bins of equal number of particles, 
from 100 to 1500, and the parameters of the fitted profile did not 
changed significantly ($\lta 1$\%) in comparison to the logarithmically spaced
grid. We consider this provides a good degree of confidence of the 
results presented for our fits.

%%%%%%%%%%%%%%%%%%%%%%%%%
\begin{table}
 \centering
  \caption{Initial Galaxy Slopes}\label{tab:glxmodels2}
  \begin{tabular}{cccc}  \hline\hline
   Merger    &  \multicolumn{3}{c}{$N$-body Model}   \\
  $M_{\rm s}/M_{\rm p}$  &{\sc rms} &  {\sc rms}$_{10}$ & $-\beta_{\rm c}$ \\
 \hline
 $S1$ & 0.098 & 0.046 & 1.25    \\
  & 0.115 & 0.068 & 1.22  \\ \hline
 $S2$  & 0.128 & 0.105 & 1.28 \\
    &  0.178 & 0.120 & 1.06  \\ \hline
 $S3$ & 0.102 & 0.075 & 1.12   \\
   & 0.100 & 0.053 & 1.32 \\
\hline\hline
\end{tabular} 
\end{table}
%%%%%%%%%%%%%%%%%%%%%%%%%%%%%%%

The mean {\sc rms} of the deviations,
$\delta=(\rho - \rho_{*})/\rho$, for the initial profiles 
in the complete fitting interval is
$0.120$, while for the inner region up to a radius of 10\% the
virial radius a mean, {\sc rms}$_{10}$, a value of $0.078$ is obtained. 
The latter value is an indication that the $N$-body realization follows
closely a NFW profile at the inner region.

 We estimated the logarithmic derivative at $r_{\rm c}$, $\beta_{\rm c}$, 
for each individual galaxy, and a mean value of $-1.21$ is 
found. Individual values of $\beta_{\rm c}$ and {\sc rms}, both 
total and inside $0.10R_{\rm v}$, are listed in 
Table~\ref{tab:glxmodels2}. We note that the values of $\beta_{\rm c}$ 
differ from each other since  $r_{\rm c}$  for each progenitor is different.

%%%%%%%%%%%%%%%%%%%%%%%%%%%%%%%%%%
\section[]{Merger Density Profiles}
%%%%%%%%%%%%%%%%%%%%%%%%%%%%%%%%%%%

%%%%%%%%%
\begin{figure}[t]
\centering
\includegraphics[width=8.3cm]{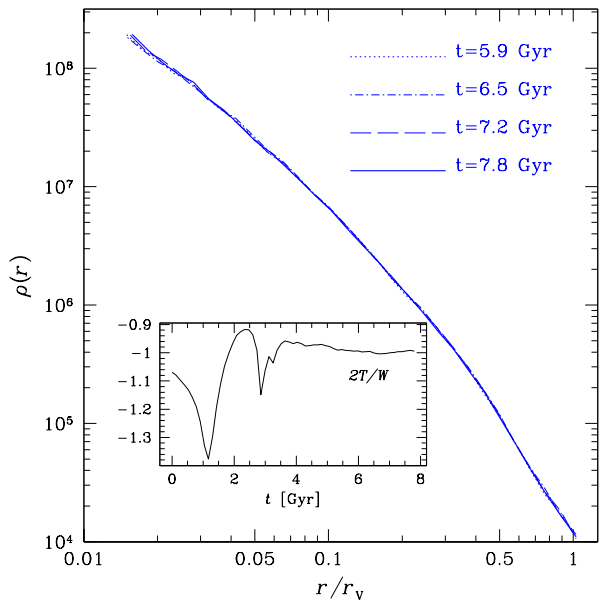}
\vspace{-0.7cm}
\caption{Late time evolution of the density profile of the remnant
of simulation $S2$, from $\approx 6\,$Gyr to  $\approx 8\,$Gyr. The 
dynamical time of the remnant is $t_{\rm dyn}\approx 0.4\,$Gyr.
The inset shows the virial ratio  during the
total time of the simulation.}
\label{fig:virialevol}
\end{figure}
%%%%%%%%%%%

In Figure~\ref{fig:virialevol} we show the time evolution of the 
dark matter density profile for merger $S2$, from 
 $\approx 6\,$Gyr to  $\approx 8\,$Gyr. As observed, the inner density
profile does not change significantly from  $\approx 6\,$Gyr to the
end of the simulation. The evolution of the virial ratio
  for the whole time of the simulation is shown in the inset.
 The dynamical time of the remnant is $t_{\rm dyn}\approx 0.4\,$Gyr, so
for $\approx  5 t_{\rm dyn}$ the profile remains unchanged  
 and the virial ratio has settled to the equilibrium value around 1; with 
some numerical fluctuations less than $1$\%. 
 A similar behavior in the profile and
virial ratio was observed for the other
two simulations considered. 
 The analysis of the resulting profile of the remnants
 were done at $t\approx 8\,$Gyr.

A NFW profile was fitted 
to the resulting merger remnants. The fitting is done in 
their corresponding interval ($r_{\rm c},r_{\rm v}$). 
In Table~\ref{tab:results} we present the values for the total {\sc rms}
 deviation of the fit (column 2), the {\sc rms} 
inside $0.1r_{\rm v}$  (column 3), the scale radius 
${\bar r}_{\rm s}$ in kpc (column 4), the concentration parameter 
${\bar c}=r_{\rm v}/{\bar r}_s$ (column 5), and the logarithmic derivative 
at $r_{\rm c}$  obtained from the fitting (column 6).

\begin{figure}[!t]
\centering
\includegraphics[width=8cm]{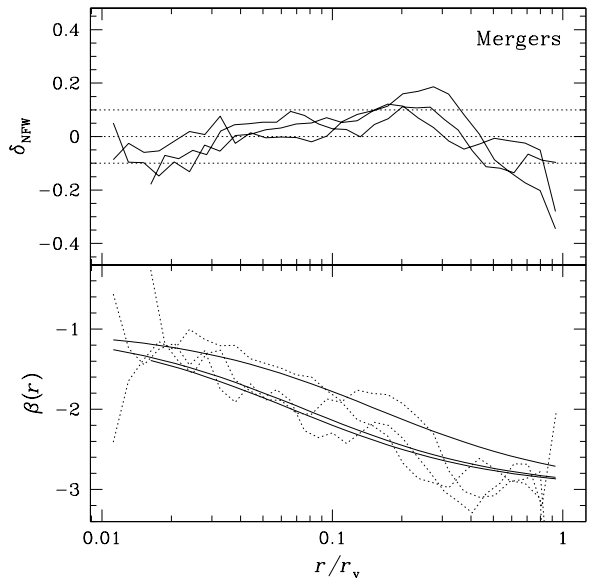}
\vspace{-0.7cm}
\caption{ ({\it Top}) Fractional residuals, 
$\delta_{\rm NFW}=(\rho - \rho_{\rm NFW})/\rho$, of the mergers'  
 density profiles with respect a NFW profile. ({\it Bottom}) 
Numerical logarithmic derivatives (dotted lines) and the ones obtained 
from the fitted profile (continuous lines). }
\label{fig:residuals}
\end{figure}

In Figure~\ref{fig:residuals} ({\it top}) we show the fractional residuals 
of the 
merger fittings to a NFW profile
and  ({\it bottom}) the numerical logarithmic derivatives 
of the data that correspond to the fitted profile (\ref{eq:nfw}). 
Though the numerical logarithmic derivatives are rather noisy, especially
at the boundaries of the fitting region, they follow in average the
$\beta(r)$ obtained from the fits. 
 The mean {\sc rms} of the density fits for the entire interval 
is $0.093$, while for the region from $r_{\rm c}$ up to 
$0.10 r_{\rm v}$ is about $0.066$. On  other hand, the average value 
of $\beta_{\rm c}$ obtained for the remnants is $-1.24$.

%%%%%%%%%%%%%%%%%%%%%%%%%
\begin{table}[!t]
 \centering
  \caption{Remnants' NFW profile.}\label{tab:results}
  \begin{tabular}{lccrrc}
  \hline\hline
   Merger & \multicolumn{5}{c}{Profile Parameters}  \\ \hline
    & {\sc rms} &  {\sc rms}$_{10}$ & $\bar{r}_{\rm s}$ & 
${\bar c}$  & $-\beta_{\rm c}$   \\
 \hline
$S1$ & 0.074 & 0.040 &   5.69 & 13.17 & 1.24 \\
$S2$ & 0.124 & 0.078 &  24.40 &  6.39 & 1.12 \\
$S3$ & 0.082 & 0.081 &   4.42 & 14.96 & 1.37 \\
\hline\hline
\end{tabular} 
\end{table}
%%%%%%%%%%%%%%%%%%%%%%%%%%%%%%%5

%%%%%%%%%
%%%%%%%%%%%

The mean  {\sc rms}$_{10}$ in the fits of the remnants is about the same
as that found in the fit for the progenitors to the NFW profile. The
values of $\beta_{\rm c}$ of the remnants (Table~3) do not necessarily
 need to coincide with, say, the average value of their progenitors
 (Table~2) since $r_{\rm c}$ is different in both cases. 
We consider  the  {\sc rms}$_{10}$ of these fits to be a better 
indicator of the preservation of the initial
cuspy profile at the inner region.

\begin{figure}
\centering
\includegraphics[width=8cm]{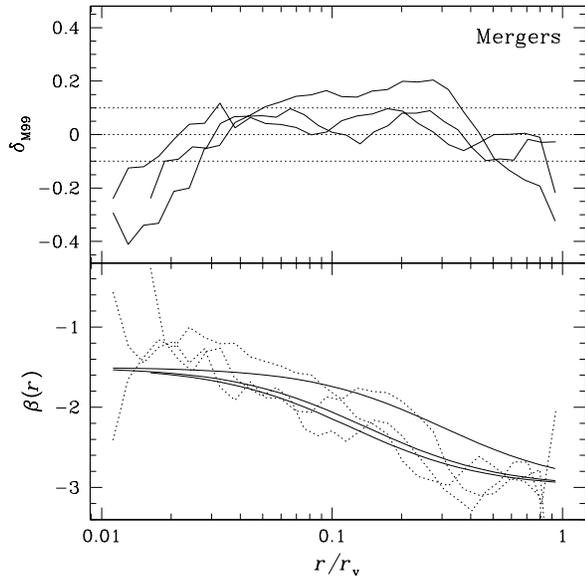}
\vspace{-0.7cm}
\caption{ ({\it Top}) Fractional residuals, $\delta_{M99}=(\rho - \rho_{\rm
    M99})/\rho$, of the mergers' density profiles with respect a M99 profile. ({\it Bottom}) numerical logarithmic derivatives (dotted lines) and the ones obtained from the assumed profile (continuous lines). }
\label{fig:residualsM99}
\end{figure}

There has been some discussion (e.g., Navarro et al. 2004) if a profile of 
the form proposed by M99,
\begin{equation}\label{eq:moore}
\rho(r) = \frac{ \rho_{\rm M} }{ (r/r_{\rm M})^{1.5} [1 + (r/ 
r_{\rm M})^{1.5} ] } \;,
\end{equation}
provides a better fit to the dark halos. We tested also if 
 (\ref{eq:moore}) provided a better fit to the 
density profiles of the remnants. 
In Table~\ref{tab:moore} we list
the values obtained for similar quantities as shown in Table~\ref{tab:results};
but with ${\tilde c} = r_{\rm v}/r_{\rm M}$. In Figure~\ref{fig:residualsM99}
 we show the fractional residuals of the fits and $\beta(r)$ using 
(\ref{eq:moore}). 

 We find a mean $\beta_{\rm c}=-1.53$ but with the average 
{\sc rms}$_{10}$ to be $0.137$. A value
 that is $\approx 100$\% larger than that obtained with the NFW profile 
inside $0.1R_{\rm v}$. 
We take this result as an indication that the M99 profile does not provides
a good fit to our merger remnants.

%%%%%%%%%%%%%%%%%%%%%%%%%
\begin{table}
 \centering
  \caption{Remnants' M99 profile.}\label{tab:moore}
  \begin{tabular}{lccrrc}
  \hline
   Merger & \multicolumn{5}{c}{Profile Parameters}  \\
    & {\sc rms} &  {\sc rms}$_{10}$ & $r_{\rm M}$ & 
${\tilde c}$  & $-\beta_{\rm c}$   \\
 \hline
$S1$ &  0.084 & 0.093 &  10.56 &  7.09 & 1.53 \\
$S2$ &  0.196 & 0.219 &  47.03 &  3.32 & 1.51 \\
$S3$ & 0.077 & 0.099 &   7.98 &  8.29 & 1.56 \\
\hline\hline
\end{tabular}
\end{table}
%%%%%%%%%%%%%%%%%%%%%%%%%%%%%%%5

%%%%%%%%%
%%%%%%%%%%%

%%%%%%%%%%%%%%%%%%%%%%
\section{Summary}
%%%%%%%%%%%%%%%%%%%%%%%

We have built up disk galaxies using the cosmologically motivated
model of MMW to study the cuspyness properties of merger remnants. 
The progenitor galaxies have spin in their halos and mass-ratios of 
about 1:10, 1:3, and 1:1. This work was restricted to study binary mergers 
following parabolic encounters.

Our results indicate that angular momentum, both intrinsic and orbital, do
not change the behavior of the density profile inside the inner regions of 
a merger remnant.

However, taking into account that few simulations were done,  
we can not completely rule out the effect of the initial 
angular momentum in the system on the inner density profile of the remnants. A 
more wider set of simulations is required to check the result found here.

%%%%%%%%%%%%%%%%%%%%%%%%%%%%%
\section*{Acknowledgments}
%%%%%%%%%%%%%%%%%%%%%%%%%%%%

We thank {\sc CONACyT}, M\'exico, for financial support through
 Project 37506-E. An anonymous referee is thanked for providing important 
comments that help to improve the presentation of this work.

%%%%%%%%%%%%%%%%%%%%%%%%%%%

%%%%%%%%%%%%%55

\end{document}